\documentclass[aps,twocolumn,groupedaddress,showpacs]{revtex4}
\usepackage[dvips]{graphicx}

\begin{document}


\title{Onsager coefficients of a Brownian Carnot cycle} 


\author{Yuki Izumida}
\email[]{izumida@statphys.sci.hokudai.ac.jp}
\author{Koji Okuda}
\email[]{okuda@statphys.sci.hokudai.ac.jp}
\affiliation{Division of Physics, Hokkaido University, Sapporo 060-0810, Japan}

\begin{abstract}
We study a Brownian Carnot cycle introduced by T. Schmiedl
and U. Seifert [Europhys. Lett. \textbf{81}, 20003 (2008)] from a viewpoint of the linear irreversible thermodynamics. 
By considering the entropy production rate of this cycle,  
we can determine thermodynamic forces and fluxes of the cycle and calculate
the Onsager coefficients
for general protocols, that is, arbitrary schedules 
to change the potential confining the Brownian particle.
We show that these Onsager coefficients contain the information of the 
protocol shape
and they satisfy the tight-coupling  
condition irrespective of whatever protocol shape
we choose. These properties may give an explanation why the
Curzon-Ahlborn efficiency often appears in the finite-time heat engines. 
\end{abstract} 

\pacs{05.70.Ln}

\maketitle

\section{Introduction}
Thermodynamics has been developed from the analysis of heat engines.                             
Carnot invented an idealized mathematical model of heat engines,    
now called the Carnot cycle, and proved that there exists a fundamental 
upper bound of the efficiency of all heat engines, which
is given by the Carnot efficiency
$\eta_\mathrm{C}\equiv 1-T_\mathrm{c}/T_\mathrm{h}$, where $T_\mathrm{h}$ and
$T_\mathrm{c}$ are the temperatures of the hotter and the colder heat 
reservoirs, respectively. 
But to attain the upper bound, we need to operate the heat engines
infinitely slowly (quasistatic limit) not to cause irreversibility. In
the quasistatic limit, the heat engine is of no practical use, because
the power, defined as work output per unit time, becomes $0$.

Practically we need to operate the heat engines in a finite
time to obtain a finite power. Curzon and Ahlborn~\cite{CA} considered 
a phenomenological model of a finite-time Carnot heat engine
under the assumptions that heat flow obeys the linear Fourier law and that irreversibility occurs only due to
the heat flow (endoreversible approximation) (see also \cite{C,N}),
and they derived that
the efficiency at the maximal power $\eta_\mathrm{max}$ of that engine becomes
\begin{eqnarray}
&&\eta_\mathrm{max}=1-\sqrt{\frac{T_\mathrm{c}}{T_\mathrm{h}}}\equiv
 \eta_\mathrm{CA}.\label{eq.1}\\
&&\text{(Curzon-Ahlborn (CA) efficiency)}\nonumber
\end{eqnarray} 
Although their derivation of $\eta_\mathrm{CA}$ may seem model-specific, they suggested that
$\eta_\mathrm{CA}$ closely approximates $\eta_\mathrm{max}$ in a few
examples of real heat engines~\cite{CA} and, above all, the simple form Eq.~(\ref{eq.1}) of  
$\eta_\mathrm{CA}$ implied its universality. In fact,  
subsequent various theoretical studies indeed revealed some sort of universality of the CA
efficiency~\cite{R1,R2,LL,G,K,B,AB,AB2,VB,CH,CH2,GS,BK,BB,TYJ,REC,B2,IO,IO2,IO3,SS,TU,ELB,ELB2,BJM,AJM,J,EKLB,ZS}. 

Recently Van den Broeck addressed the generality of the CA efficiency
from a viewpoint of
the linear irreversible thermodynamics~\cite{VB}. He described heat engines by 
using the Onsager relations 
\begin{eqnarray}
&&J_1=L_{11}X_1+L_{12}X_2,\label{eq.2}\\
&&J_2=L_{21}X_1+L_{22}X_2,\label{eq.3}
\end{eqnarray} 
and proved that $\eta_\mathrm{CA}$ is the upper bound of the
efficiency at the maximal power $\eta_\mathrm{max}$ in the linear response regime $\Delta T
\to 0$, where $\Delta T\equiv T_\mathrm{h}-T_\mathrm{c}$. This upper
bound can be attained when the Onsager coefficients $L_{ij}$'s 
in Eqs.~(\ref{eq.2}) and (\ref{eq.3}) satisfy 
the tight coupling condition $|q|=1$, where $q\equiv L_{12}/\sqrt{L_{11}L_{22}}$ is called the coupling 
strength parameter. 
Various theoretical models of the heat engines can be understood by using the
Onsager relations, ranging from steady state Brownian 
motors~\cite{GS,BK,B2,BB,TYJ,REC} to a 
macroscopic finite-time Carnot cycle~\cite{IO3}. 

Recently Schmiedl and Seifert suggested an analytically tractable model
of a Carnot heat engine from which useful work can be extracted through the motion
of a Brownian particle by changing the 
potential confining the particle~\cite{SS}. 
In this paper, we call their model a Brownian Carnot cycle.
By restricting the potential to the harmonic form, they analytically
showed that the efficiency at the maximal power $\eta_\mathrm{max}$
agrees with $\eta_\mathrm{CA}$ in the limit of $\Delta T \to 0$. (See
Sec.~II for details.) The remarkable feature of their analysis is that 
they compared all protocols when maximizing the power, where 
we mean by {\it protocol} the schedule to change the potential.

Although they showed that their $\eta_\mathrm{max}$ attains
$\eta_\mathrm{CA}$ 
in the linear response regime $\Delta T \to 0$, 
but the explicit Onsager relations Eqs.~(\ref{eq.2}) and
(\ref{eq.3}) have not been obtained yet, especially how the information
of the protocol is reflected in the Onsager relations is unclear. 
In this paper, we calculate the Onsager coefficients $L_{ij}$'s of the
Brownian Carnot cycle with the harmonic potential
for general protocols, that is, arbitrary schedules to change the
potential. 

The organization of this paper is as follows. In Sec.~II, we introduce
the Brownian Carnot cycle and review its general properties according to~\cite{SS}. 
We demonstrate the calculations of its 
Onsager coefficients in Sec.~III and
discuss physical implications of them in Sec.~IV.   
We summarize this study in Sec.~V.    
\section{Model}
\begin{figure}
\begin{center} 
\includegraphics[scale=0.275]{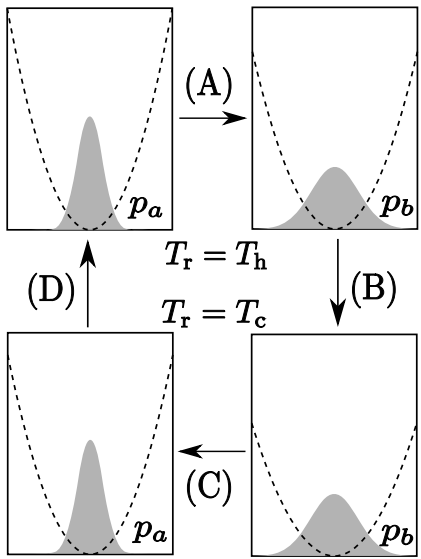}  
\end{center}   
\caption{Schematic illustration of a Brownian Carnot cycle.
A Brownian particle immersed in noisy thermal surroundings at
temperature $T_\mathrm{r}$
is operated by changing the spring constant of the harmonic potential
(dashed lines). 
The shadow regions represent the probability distribution $p(x,t)$.
(A) and (C) are isothermal steps at $T_\mathrm{r}=T_\mathrm{h}$ and at $T_\mathrm{r}=T_\mathrm{c}$, respectively.
(B) and (D) are instantaneous adiabatic steps which do not generate irreversibility.
During the isothermal step (A) ((C)),
$p(x,t)$ evolves according to the 
Fokker-Planck equation from $p_a$ ($p_b$) to $p_b$ ($p_a$).
}\label{fig.1}
\end{figure}     
Let us introduce the Brownian Carnot cycle~\cite{SS}.      
Imagine that a one-dimensional overdamped 
Brownian particle is immersed in thermal surroundings at a
temperature $T_\mathrm{r}$ ($\mathrm{r}=\mathrm{h}, \mathrm{c}$) and is trapped in a harmonic potential. 
The surroundings behave like noise on the particle and fluctuate its position.
Therefore its motion becomes probabilistic rather than deterministic. 
In this situation, the time evolution of the probability distribution $p(x,t)$
of the position $x$ of the particle at time $t$ can be described by the Fokker-Planck equation
\begin{eqnarray} 
\frac{\partial p(x,t)}{\partial t}=-\mu \frac{\partial}{\partial x} \cdot \biggl[-\frac{\partial}{\partial
 x} V(x,t)-T_\mathrm{r}\frac{\partial}{\partial 
 x}\biggr]p(x,t),\label{eq.4} 
\end{eqnarray} 
where $V(x,t)$ is the potential energy, which is given by $V(x,t)=\lambda (t)x^2/2$,
using the time-dependent spring constant $\lambda(t)$. $\mu$ is the
mobility of the particle and we set the Boltzmann constant $k_\mathrm{B}$ as unity.
We can construct a Brownian Carnot cycle in a statistical sense by using Eq.~(\ref{eq.4}) as
follows (see Fig.~\ref{fig.1}): (A). Isothermal step at 
$T_\mathrm{r}=T_\mathrm{h}$ ($0 < t < t_{A}$): by changing the spring
constant, we change the
distribution function from initial $p_a(x)$ to $p_b(x)$ for the duration of  
$t_{A}$. (B). Adiabatic step: at time $t=t_{A}$, we   
switch the reservoir at $T_\mathrm{h}$   
to a colder one at $T_\mathrm{c}$ instantaneously. Because this
instantaneous adiabatic step gives the probability distribution no time
to relax, the probability distribution remains $p=p_b(x)$ during this step.
(C). Isothermal step at $T_\mathrm{r}=T_\mathrm{c}$ ($t_{A} < t < t_{A}+t_{C}$): by 
changing the spring constant, we change the 
distribution function from $p_b(x)$ to $p_a(x)$ for the duration of 
$t_{C}$. (D). Adiabatic step: at time $t=t_{C}$, we switch the reservoir at $T_\mathrm{c}$ to the one at
$T_\mathrm{h}$ instantaneously again. This adiabatic step also keeps the probability
distribution as $p=p_a(x)$ and then the cycle closes. Since $\lambda(t)$
determines the schedule to change the potential, we consider
$\lambda(t)$ as the protocol in this Brownian Carnot cycle.   
During the isothermal steps (A) and (C), the probability distribution
evolves according to the Fokker-Planck equation Eq.~(\ref{eq.4}) and we also assume that
the initial probability $p_a(x)$ is Gaussian with the mean $0$ and the variance $w_a$ as
$p_a(x)=\exp(-x^2/(2w_a))/\sqrt{2\pi w_a}$. Then it can be shown that
the probability distribution always
remains Gaussian with the mean $0$ in
the case of the harmonic potential if we initially prepare it so. 

Here we define the internal energy $E(t)$ and the entropy $S(t)$ of the
particle at time $t$ as 
\begin{eqnarray}  
E(t)\equiv \int_{-\infty}^{\infty} dx\ p(x,t)V(x,t),\label{eq.5}\\ 
S(t)\equiv -\int_{-\infty}^{\infty} dx\ p(x,t)\ln p(x,t),\label{eq.6}
\end{eqnarray} 
respectively.
The entropy $S$ of the      
particle does not change during the    
adiabatic steps (B) and (D) because the probability distribution remains unchanged.
Thus the adiabatic steps are surely isentropic, 
which do not generate irreversibility during them at all.    

To calculate the work output during the cycle, we consider the 
time evolution equation of the variance $w(t)\equiv 
\int_{-\infty}^{\infty} dx \ p(x,t)x^2$ during the isothermal steps (A)
and (C) by using Eq.~(\ref{eq.4}) as 
\begin{eqnarray} 
&&\frac{dw^{(A)}}{dt}=-2\mu \lambda w^{(A)}+2\mu T_\mathrm{h} \ (0<t<t_{A}),\label{eq.7}\\
&&\frac{dw^{(C)}}{dt}=-2\mu \lambda w^{(C)}+2\mu T_\mathrm{c} \ (t_{A}<t<t_{A}+t_{C}),\label{eq.8}
\end{eqnarray}
where we have defined the variance during the isothermal steps (A) and
(C) as $w^{(A)}$ and $w^{(C)}$, respectively.
Then the work output in the isothermal step (A) ($0 < t < t_{A}$) can be calculated
by using Eq.~(\ref{eq.7}) as
\begin{eqnarray} 
W^{(A)}&&=-\int_{0}^{t_{A}}dt \int_{-\infty}^{\infty} dx \ p(x,t)\frac{\partial V}{\partial
t}\nonumber\\ 
&&=-\int_{0}^{t_{A}}dt \  \frac{d\lambda}{dt}\frac{w^{(A)}}{2}\nonumber \\ 
&&=-\frac{1}{4\mu}\int_{0}^{t_{A}}\frac{{(\frac{dw^{(A)}}{dt})}^2}{w^{(A)}}dt+\frac{1}{2}T_\mathrm{h}[\ln
w^{(A)}]_{0}^{t_{A}}\nonumber\\
&&-\frac{1}{2}[\lambda w^{(A)}]_{0}^{t_{A}}\nonumber \\  
&&\equiv -W_\mathrm{irr}^{(A)}+T_\mathrm{h}\Delta {S}^{(A)}-\Delta {E}^{(A)},\label{eq.9}
\end{eqnarray}  
where $W_{\mathrm{irr}}^{(A)}\equiv
\frac{1}{4\mu}\int_{0}^{t_\mathrm{A}}(\frac{dw^{(A)}}{dt})^2/w^{(A)} \
dt$ is the decrease of the work by irreversibility, which vanishes in the quasistatic limit
$t_{A} \to \infty$.
Assuming that the probability distribution $p$ is always Gaussian with
the variance $w$, 
we have also defined the internal energy change and the entropy change
during the isothermal step (A) as
${\Delta
S}^{(A)}\equiv \frac{1}{2}[\ln
w^{(A)}]_{0}^{t_{A}}$ and ${\Delta
E}^{(A)}\equiv \frac{1}{2}[\lambda w^{(A)}]_{0}^{t_{A}}$,  
by using Eqs~(\ref{eq.5}) and (\ref{eq.6}), respectively.  
The work output in the isothermal step (C) ($t_{A} < t < t_{A}+t_{C}$) can
also be calculated likewise as
\begin{eqnarray} 
W^{(C)}&&=-\int_{t_{A}}^{t_{A}+t_{C}}dt \int_{-\infty}^{\infty} dx \ p(x,t)\frac{\partial V}{\partial
t}\nonumber\\
&&=-\int_{t_{A}}^{t_{A}+t_{C}}dt \  \frac{d\lambda}{dt}\frac{w^{(C)}}{2}\nonumber \\ 
&&=-\frac{1}{4\mu}\int_{t_{A}}^{t_{A}+t_{C}}\frac{{(\frac{dw^{(C)}}{dt})}^2}{w^{(C)}}dt+\frac{1}{2}T_\mathrm{c}[\ln
w]_{t_{A}}^{t_{A}+t_{C}}\nonumber \\
&&-\frac{1}{2}[\lambda w^{(C)}]_{t_{A}}^{t_{A}+t_{C}}\nonumber \\  
&&\equiv -W_\mathrm{irr}^{(C)}+T_\mathrm{c}\Delta {S}^{(C)}-\Delta {E}^{(C)},\label{eq.10}
\end{eqnarray}
where $W_{\mathrm{irr}}^{(C)}$, ${\Delta S}^{(C)}$ and ${\Delta
E}^{(C)}$ are defined in the same way as in the step (A).
We need to determine the dynamics of $w^{(i)}(t)$ ($i=A, C$) to calculate $W_\mathrm{irr}^{(A)}$
and $W_\mathrm{irr}^{(C)}$ explicitly. Since the cyclic change of
$\lambda (t)$ asymptotically leads to the Gaussian cyclic change of
$p(x,t)$, $w^{(i)}(t)$ ($i=A, C$) is uniquely determined by 
$\lambda(t)$. Reversely, if we determine $w^{(i)}(t)$ ($i=A, C$), 
then the dynamics of $\lambda(t)$ is uniquely determined 
via Eqs.~(\ref{eq.7}) and (\ref{eq.8}). Therefore we may regard
$w^{(i)}(t)$ ($i=A, C$) as the protocol in
this Brownian Carnot cycle instead of $\lambda(t)$. 
 
The work outputs $W^{(B)}$ and $W^{(D)}$ in the adiabatic steps (B) ($t=t_{A}$) and
(D) ($t=t_{A}+t_{C}$) simply
become the change of the internal energy of the system as
$W^{(B)}=-\Delta E^{(B)}$ and $W^{(D)}=-\Delta E^{(D)}$ 
because the instantaneous
change of the spring constant during the adiabatic steps does not affect
the probability distribution.
Then the total work output $W\equiv \sum_{i=A}^{D}W^{(i)}$  
during the entire cycle (A)-(D) is summed as
\begin{eqnarray}
W=-W^{(A)}_\mathrm{irr}-W^{(C)}_\mathrm{irr}+(T_\mathrm{h}-T_\mathrm{c})\Delta S,\label{eq.11}
\end{eqnarray}
where we have used relations $\sum_{i=A}^{D} \Delta E^{(i)}=0$ and $\sum_{i=A}^{D}   
\Delta S^{(i)}=0$ due to the periodicity of the system and have   
defined $\Delta S$ as
\begin{eqnarray} 
\Delta S\equiv \Delta S^{(A)}=-\Delta S^{(C)}=\ln
\sqrt{\frac{w_b}{w_a}}.\label{eq.12}
\end{eqnarray}   
The heat $Q_\mathrm{h}$ absorbed into the system from the hotter reservoir during the step (A) becomes
\begin{eqnarray}  
Q_\mathrm{h}\equiv \Delta E^{(A)}+W^{(A)}=T_\mathrm{h}\Delta S-W^{(A)}_\mathrm{irr}.\label{eq.13}
\end{eqnarray} 
Here we define the power and the efficiency as
\begin{eqnarray}
&&P\equiv \dot{W}\equiv \frac{W}{t_{A}+t_{C}},\label{eq.14}\\
&&\eta \equiv \frac{\dot{W}}{\dot{Q}_\mathrm{h}},\label{eq.15} 
\end{eqnarray}
where the dot
denotes a quantity divided by the one-cycle period or a quantity 
per unit time throughout the
paper. In the Brownian Carnot cycle, the one-cycle period is $t_{A}+t_{C}$.

After the above setup for the Brownian Carnot cycle, Schmiedl and Seifert
considered the maximization of the power as
follows. Firstly they  
calculated the optimal protocol which maximizes the power Eq.~(\ref{eq.14})
under the condition that the durations $t_{A}$ and
$t_{C}$ and the boundary values $w_a$ and $w_b$ are fixed. 
This optimization can be realized by minimizing the functionals in 
$W^{(A)}_\mathrm{irr}$ and $W^{(C)}_\mathrm{irr}$ since $\Delta S$ in
Eq.~(\ref{eq.11}) is constant. By solving Euler-Lagrange
equations for these functionals with given boundary values $w_a$ and
$w_b$, they obtained the optimal protocol as 
\begin{eqnarray}
w^{(A)}(t)&&=w_a(1+(\sqrt{w_b/w_a}-1)t/t_{A})^2,\label{eq.16}\\
w^{(C)}(t)&&=w_b(1+(\sqrt{w_a/w_b}-1)(t-t_{A})/t_{C})^2.\label{eq.17}
\end{eqnarray}
Then, $W^{(A)}_\mathrm{irr}$ and $W^{(C)}_\mathrm{irr}$
under this protocol are expressed as
\begin{eqnarray}
&&W^{(A)}_\mathrm{irr}=\frac{(\sqrt{w_b}-\sqrt{w_a})^2}{\mu}\cdot \frac{1}{t_{A}},\label{eq.18}\\
&&W^{(C)}_\mathrm{irr}=\frac{(\sqrt{w_b}-\sqrt{w_a})^2}{\mu}\cdot \frac{1}{t_{C}}. \label{eq.19}
\end{eqnarray}
Secondly they considered the further maximization of the power of the
optimal protocol by changing the durations $t_{A}$ and
$t_{C}$. From $\partial P/\partial t_{A}=\partial
P/\partial t_{C}=0$, $t_{A}$ and $t_{C}$ can be determined as
\begin{eqnarray}
t_{A}=t_{C}=\frac{4(\sqrt{w_b}-\sqrt{w_a})^2}{\mu(T_\mathrm{h}-T_\mathrm{c})\Delta S}.\label{eq.20}
\end{eqnarray}
Finally, they obtained the maximal power $P_\mathrm{max}$ and the efficiency at the maximal power $\eta_\mathrm{max}$ as
\begin{eqnarray}
&&P_\mathrm{max}=\frac{\mu\left(\ln
\sqrt{{w_b}/{w_a}}\right)^2{\Delta T}^2}{16(\sqrt{w_b}-\sqrt{w_a})^2},\label{eq.21}\\ 
&&\eta_\mathrm{max}=\frac{\eta_\mathrm{C}}{2-\eta_\mathrm{C}/2},\label{eq.22}
\end{eqnarray}
where $\eta_\mathrm{C}=1-T_\mathrm{c}/T_\mathrm{h}$ is the Carnot efficiency. This result is
remarkable because $\eta_\mathrm{max}$ is independent of the boundary
values $w_a$ and $w_b$, although $P_\mathrm{max}$ depends on them.
Thus, this implies that $\eta_\mathrm{max}$ in Eq.~(\ref{eq.22}) remains
unchanged even when the further maximization of the power is performed
by changing $w_a$ and $w_b$ in Eq.~(\ref{eq.21}). Therefore,
$\eta_\mathrm{max}$ in Eq.~(\ref{eq.22}) gives the efficiency at the
maximal power under arbitrary protocols of $w(t)$.

Moreover, by expanding $\Delta 
T$, they obtained $\eta_\mathrm{max}=\Delta T/(2T)+O({\Delta T}^2)$ which
is equal to $\eta_\mathrm{CA}$ in the linear order of $\Delta T$. 
A remark should be added here. The essential point of   
their derivation is to consider the two-step maximization of the power:
first by the optimal protocol for the fixed durations and second by 
those durations. 
This strategy to find the maximal power 
is indeed effective for the non-linear response regime of the cycle where 
we have no general theory to describe that regime at present. 
But we can use the linear irreversible thermodynamics in the linear 
order in $\Delta T$. As we will show in Sec.~III, we indeed apply the
framework of the linear irreversible thermodynamics to this Brownian Carnot cycle and can extract
more detailed information on the maximization of the power in the linear
response regime.
\section{Calculations of the Onsager coefficients}  
In this Section, we will calculate the Onsager coefficients of the
Brownian Carnot cycle.  
Firstly we introduce the Onsager relations 
of a general linear irreversible heat engine~\cite{VB}. 
Let us consider that the heat engine is working under an external
force $F$ and a small temperature difference $\Delta T$. 
The work performed against the external force $F$ is defined as $W=-Fx$,
where $x$ is the thermodynamically conjugate variable of $F$. In the limit
of the small temperature difference $\Delta T \to 0$, we can  
define a thermodynamic force $X_1=F/T_\mathrm{c}\simeq F/T$ where
$T\equiv (T_\mathrm{h}+T_\mathrm{c})/2$ and its conjugate flux as $J_1=\dot{x}$.
We also define the inverse temperature difference $1/T_\mathrm{c}-1/T_\mathrm{h}\simeq \Delta T/T^2$ as another thermodynamic
force $X_2$ and the heat flux
from the hotter heat reservoir $\dot{Q}_\mathrm{h}$ as its conjugate flux
$J_2$. Moreover, we assume that the Onsager relations
hold for these fluxes and forces as~\cite{O,GM} 
\begin{eqnarray}   
&&J_1=L_{11}X_1+L_{12}X_2,\label{eq.23}\\
&&J_2=L_{21}X_1+L_{22}X_2,\label{eq.24} 
\end{eqnarray}    
where $L_{ij}$'s are the Onsager coefficients with the symmetry relation $L_{12}=L_{21}$.

To write down the Onsager relations for the Brownian Carnot cycle,
we need to choose the thermodynamic fluxes and forces for it. 
A typical way for choosing them is to consider the total entropy production rate $\dot{\sigma}$ 
during one cycle. Because the system entropy $S$ does not change after
one cycle as $\sum_{i=A}^{D}\Delta S^{(i)}=0$, only the entropy
increase in the reservoirs contributes to $\dot{\sigma}$ as
\begin{eqnarray}
\dot{\sigma}\equiv -\frac{\dot{Q}_\mathrm{h}}{T_\mathrm{h}}-\frac{\dot{Q}_\mathrm{c}}{T_\mathrm{c}}=-\frac{\dot{Q}_\mathrm{h}}{T_\mathrm{h}}-\frac{\dot{W}-\dot{Q}_\mathrm{h}}{T_\mathrm{c}}.\label{eq.25}
\end{eqnarray}
In the linear response regime $\Delta T \to 0$, $\dot{\sigma}$ can be
approximated as  
\begin{eqnarray}
\dot{\sigma}\simeq \frac{-W}{T}\cdot
\frac{1}{(\alpha+1)t_{A}}+\frac{\Delta T}{T^2}\cdot \dot{Q}_\mathrm{h},\label{eq.26}
\end{eqnarray}
where we define $\alpha\equiv t_{C}/t_{A}$ as the ratio of $t_{C}$ to $t_{A}$ and
we have neglected the higher terms like $O(\Delta T \dot{W})$ and
$O(\Delta T^3\dot{Q}_\mathrm{h})$, whose reason will be clarified later. 
The expression $\dot{\sigma}=J_1X_1+J_2X_2$ by the linear irreversible thermodynamics
leads to the decomposition of the thermodynamic forces as 
\begin{eqnarray} 
X_1\equiv \frac{-W}{T},\ \label{eq.27}
X_2\equiv \frac{\Delta T}{T^2}
\end{eqnarray}
and their conjugate fluxes as
\begin{eqnarray}
J_1\equiv \frac{1}{(\alpha+1)t_{A}}, \ J_2\equiv \dot{Q}_\mathrm{h}.\label{eq.28}
\end{eqnarray}
As is clear from these definitions for the thermodynamic fluxes and
forces, we understand that the higher terms like $O(\Delta T \dot{W})$ and
$O(\Delta T^3\dot{Q}_\mathrm{h})$ do not contribute to
the entropy production rate $\dot{\sigma}$ in the linear response regime $\Delta T \to 0$. That's why we
neglected them in Eq.~(\ref{eq.26}). 
By using these definitions for the thermodynamic fluxes and forces, we can write down the Onsager relations of the
Brownian Carnot cycle as 
\begin{eqnarray}
\frac{1}{(\alpha+1)t_{A}}&&=L_{11}\frac{-W}{T}+L_{12}\frac{\Delta T}{T^2},\label{eq.29}\\
\dot{Q}_\mathrm{h}&&=L_{21}\frac{-W}{T}+L_{22}\frac{\Delta T}{T^2}.\label{eq.30}
\end{eqnarray}

Now we calculate the Onsager coefficients $L_{ij}$'s.
First we calculate $L_{11}$ and $L_{21}$ as follows. 
By using a scaled variable $s\equiv
t/t_{A}$ $(0 < s < 1)$ instead of $t$ ($0 < t < t_{A}$) during the isothermal step (A), 
$W_\mathrm{irr}^{(A)}$ in Eq. (\ref{eq.11}) becomes   
\begin{eqnarray}  
W_\mathrm{irr}^{(A)}=\frac{1}{t_{A}}\frac{1}{4\mu}\int_{0}^{1}\frac{(\frac{d\tilde{w}^{(A)}}{ds})^2}{\tilde{w}^{(A)}}ds\equiv
 \frac{A_\mathrm{irr}}{t_{A}},\label{eq.31} 
\end{eqnarray}  
where we call $\tilde{w}^{(A)}(s)\equiv w^{(A)}(t)$ the protocol shape. 
In Eq.~(\ref{eq.31}), we have divided $W_\mathrm{irr}^{(A)}$ into the part proportional
to the functional of the protocol shape $A_\mathrm{irr}$ and
the duration $t_A$.   
$W_\mathrm{irr}^{(C)}$ during the isothermal step (C) in
Eq.~(\ref{eq.11}) can also be divided as 
\begin{eqnarray}  
W_\mathrm{irr}^{(C)}=\frac{1}{t_{C}}\frac{1}{4\mu}\int_{0}^{1}\frac{(\frac{d\tilde{w}^{(C)}}{ds})^2}{\tilde{w}^{(C)}}ds\equiv
 \frac{C_\mathrm{irr}}{t_{C}}\label{eq.32} 
\end{eqnarray}  
by using a scaled variable
$s\equiv (t-t_{A})/t_{C}$ ($0 < s < 1$) instead of $t$ ($t_{A} < t < t_{A}+t_{C}$)
and $\tilde{w}^{(C)}(s)\equiv w^{(C)}(t)$.
Then Eq.~(\ref{eq.11}) can be expressed by using Eqs.~(\ref{eq.31}) and (\ref{eq.32}) as
\begin{eqnarray}
W=-\frac{1}{t_{A}}\Bigl(A_\mathrm{irr}+\frac{C_\mathrm{irr}}{\alpha}\Bigr)+\Delta
 T\Delta S\label{eq.33}.
\end{eqnarray}
By putting 
$\Delta T=0$ in Eq.~(\ref{eq.33}), we can obtain the relation
\begin{eqnarray} 
\frac{1}{(\alpha+1)t_{A}}=\frac{T}{\left(A_\mathrm{irr}+C_\mathrm{irr}/{\alpha}\right)(\alpha+1)}\cdot
\frac{-W}{T},\label{eq.34}
\end{eqnarray}
which determines the coefficient
$L_{11}$ as   
\begin{eqnarray}
L_{11}=\frac{T}{\left(A_\mathrm{irr}+C_\mathrm{irr}/{\alpha}\right)(\alpha+1)}.\label{eq.35}
\end{eqnarray}
$\dot{Q}_\mathrm{h}$ at $\Delta T=0$ can also be calculated by using
Eqs.~(\ref{eq.13}) and (\ref{eq.34}) as 
\begin{eqnarray} 
\dot{Q}_\mathrm{h}=\frac{T^2\Delta
S}{\left(A_\mathrm{irr}+C_\mathrm{irr}/{\alpha}\right)(\alpha+1)}\cdot
\frac{-W}{T}-\frac{A_\mathrm{irr}}{(\alpha+1){t_{A}}^2},\label{eq.36}
\end{eqnarray}
where the second term can be neglected because it is $O(W^2)$ quantity from Eq.~(\ref{eq.34}). Then
the coefficient $L_{21}$ can be determined as
\begin{eqnarray}
L_{21}=\frac{T^2\Delta S}{\left(A_\mathrm{irr}+C_\mathrm{irr}/{\alpha}\right)(\alpha+1)}.\label{eq.37}
\end{eqnarray}
Next we calculate $L_{12}$ and $L_{22}$ likewise. Putting $W=0$ in Eq.~(\ref{eq.33}),
we obtain the relation
\begin{eqnarray}
\frac{1}{(\alpha+1)t_{A}}=\frac{T^2\Delta
 S}{\left(A_\mathrm{irr}+C_\mathrm{irr}/{\alpha}\right)(\alpha+1)}
\cdot \frac{\Delta T}{T^2},\label{eq.38}
\end{eqnarray}
which determines the coefficient $L_{12}$ as
\begin{eqnarray}
L_{12}=\frac{T^2\Delta S}{\left(A_\mathrm{irr}+C_\mathrm{irr}/{\alpha}\right)(\alpha+1)}.\label{eq.39}
\end{eqnarray}
From Eqs.~(\ref{eq.37}) and (\ref{eq.39}), we find that the Onsager
symmetry relation $L_{21}=L_{12}$ surely holds as expected. 
$\dot{Q}_\mathrm{h}$ at $W=0$ thus becomes 
\begin{eqnarray}
\dot{Q}_\mathrm{h}=\frac{T^3{\Delta 
 S}^2}{\left(A_\mathrm{irr}+C_\mathrm{irr}/{\alpha}\right)(\alpha+1)}\cdot
 \frac{\Delta T}{T^2}-\frac{A_\mathrm{irr}}{(\alpha+1){t_{A}}^2}\label{eq.40}
\end{eqnarray}
from Eqs.~(\ref{eq.13}) and (\ref{eq.38}), where the second term can be
neglected because it is
$O(\Delta T^2)$ quantity from Eq.~(\ref{eq.38}). 
Then the last coefficient $L_{22}$ turns out to be
\begin{eqnarray}
L_{22}=\frac{T^3{\Delta S}^2}{\left(A_\mathrm{irr}+C_\mathrm{irr}/{\alpha}\right)(\alpha+1)}.\label{eq.41}
\end{eqnarray}
Note that these Onsager
coefficients satisfy the constraints $L_{11}\ge 0$, $L_{22}\ge 0$ and
$L_{11}L_{22}-L_{12}L_{21}\ge 0$ which come from the positivity of the
entropy production rate $\dot{\sigma}$. We also note that 
although they are expressed in
terms of $\tilde{w}$'s through $A_\mathrm{irr}$ and
$C_\mathrm{irr}$, we could easily switch to the $\lambda$ representation
by solving the differential equation Eqs.~(\ref{eq.7}) and (\ref{eq.8}) explicitly and substituting the solutions $\tilde{w}$'s
into $A_\mathrm{irr}$ and $C_\mathrm{irr}$.
\section{Discussion} 
Now that we have calculated the Onsager coefficients of the Brownian Carnot  
cycle in Sec.~III, we will discuss physical implications of
them in this Section. 
To see how the Onsager relations are used to describe the efficiency and
the power of the Brownian Carnot cycle, we introduce the general theory to
describe the heat engines governed by the Onsager relations developed
in~\cite{VB} as follows.   
Generally the power $P$ and the efficiency $\eta$ of the linear
irreversible heat engine are expressed as
\begin{eqnarray} 
P&&=\dot{W}=-F\dot{x}=-J_1X_1T,\label{eq.42}\\
\eta&&=\frac{\dot{W}}{\dot{Q}_\mathrm{h}}=-\frac{J_1X_1T}{J_2},\label{eq.43}
\end{eqnarray}
respectively.
When $X_2$ determined by the reservoir's temperatures and the Onsager
coefficients $L_{ij}$'s are given, we can see that only $X_1$ determines the power
$P$ and the efficiency $\eta$. Since $X_1$ at the maximal power is given
by $\partial
P/\partial X_1=0$, we obtain $\eta_\mathrm{max}$ as
\begin{eqnarray}
\eta_\mathrm{max}=\frac{\Delta T}{2T}\frac{q^2}{2-q^2},\label{eq.44}
\end{eqnarray} 
where $q$ is defined as   
\begin{eqnarray}
q\equiv \frac{L_{12}}{\sqrt{L_{11}L_{22}}},\label{eq.45} 
\end{eqnarray}
which is called the coupling strength parameter. Taking into account
the restriction 
$-1 \le q \le 1$ due to $L_{11}L_{22}-L_{12}L_{21}\ge 0$ which comes from the positivity of the entropy
production rate $\dot{\sigma}$, $\eta_\mathrm{max}$ becomes the upper
bound
\begin{eqnarray}
\eta_\mathrm{max}=\frac{\Delta T}{2T}=\eta_\mathrm{CA}+O(\Delta T^2)\label{eq.46}
\end{eqnarray}
when
the tight coupling condition $|q|=1$ is satisfied.  

As seen from Eqs.~(\ref{eq.31}) and (\ref{eq.32}), we have shown that when we consider a protocol
$w^{(i)}(t)$ by separating it into its shape
$\tilde{w}^{(i)}(s)$ and its duration
$t_i$, the Onsager coefficients Eqs.~(\ref{eq.35}), (\ref{eq.37}),
(\ref{eq.39}) and (\ref{eq.41}) of the Brownian Carnot cycle contain the information of
the protocol shape through $A_\mathrm{irr}$ and $C_\mathrm{irr}$ and the thermodynamic flux
$J_1$ is the inverse of the one-cycle period $(\alpha+1)t_{A}=t_{A}+t_{C}$.
The remarkable feature is that as easily confirmed, they 
satisfy the
tight coupling condition 
$|q|=1$, independently of the protocol shape, which 
means that the efficiency at the maximal power is always the CA 
efficiency. As we have shown above, this result is derived when the
power is maximized by changing the thermodynamic force $X_1$.
However, $X_1$ corresponds to $J_1$ via Eq.~(\ref{eq.29}) at the fixed temperatures
and protocol shape.
Therefore, we can consider that changing $X_1$ is equivalent to changing
$J_1$ or the one-cycle period.
This leads to the result that in the present model of the Brownian
Carnot cycle, the efficiency at the maximal power is always the CA
efficiency when the power is maximized by changing only the one-cycle
period with the protocol shape fixed.
If the above feature of the Onsager coefficients is common to various
heat engines, our result may explain the appearance of the CA efficiency
found in ~\cite{CA,IO,IO2,IO3}, where the power is maximized by
changing only the one-cycle period without choosing the
optimal protocol shape realizing the true maximal power.

The relation to the study by Schmiedl and Seifert~\cite{SS} reviewed in
Sec.~II can be understood as follows. 
Though in Eq.~(\ref{eq.46}), we have seen that the efficiency at the maximal power
$\eta_\mathrm{max}$ under an arbitrary protocol shape is $\eta_\mathrm{CA}$,
the maximal power $P_\mathrm{max}$ should be lower than the true maximal
power on the space of all protocol shapes. Here let us consider the further
maximization of $P_\mathrm{max}$ obtained from the Onsager relations.
Maximizing the power Eq.~(\ref{eq.42}) by changing $X_1$, the maximal
power $P_\mathrm{max}$ is given by
\begin{eqnarray}
P_\mathrm{max}=\frac{L_{22}{\Delta T}^2}{4T^3}.\label{eq.47}
\end{eqnarray} 
We notice that $L_{22}$ still depends on $\alpha$, $A_\mathrm{irr}$ and $C_\mathrm{irr}$. 
Then we can further maximize this $P_\mathrm{max}$ by minimizing
the functionals in $A_\mathrm{irr}$ and $C_\mathrm{irr}$ with the
boundary values $w_a$ and $w_b$, and also by changing $\alpha$, which reduces to an inequality
\begin{eqnarray}
P_\mathrm{max}\le \frac{\mu\left(\ln
\sqrt{{w_b}/{w_a}}\right)^2{\Delta
T}^2}{16(\sqrt{w_b}-\sqrt{w_a})^2}.\label{eq.48} 
\end{eqnarray} 
The equality is realized when $\alpha=1$ and 
$A_\mathrm{irr}=C_\mathrm{irr}=(\sqrt{w_b}-\sqrt{w_a})^2/\mu$,
which corresponds to the case of the optimal protocol in Eqs.~(\ref{eq.16}) and (\ref{eq.17}).
The right-hand side of Eq.~(\ref{eq.48}) is the maximal power that the cycle can output in the linear
response regime at given boundary values $w_a$ and $w_b$ and is equal to Eq.~(\ref{eq.21}).
Thus the result in~\cite{SS} can be reproduced from the Onsager relations.  
\section{summary}
In summary, we derived the Onsager relations for a Brownian Carnot cycle 
working in the linear-response regime.
We considered a protocol by separating it into its shape and its
duration, where we mean by protocol the schedule to change the
potential confining the Brownian particle. Then,  
we found that the Onsager coefficients contain  
the functionals of the protocol shape to change the 
potential and they satisfy the tight 
coupling condition, irrespective of whatever protocol shape we choose. 
This result implies that we can attain the Curzon-Ahlborn
efficiency when maximizing the power by changing only the one-cycle period under an arbitrary
protocol shape. 
In this sense, the Curzon-Ahlborn efficiency looks like the
Carnot efficiency because the latter can also be attained irrespective of
whatever protocol shape we choose in the quasistatic limit.  
Although we used the harmonic potential to construct the Brownian Carnot cycle,
we believe that our results could also be applied to the models with general potentials
and arbitrary dimensions discussed in~\cite{SS}, where the efficiency at the true maximal
power agrees with the Curzon-Ahlborn efficiency.  
We expect that our study will stimulate the further discussion on the
physics of the finite-time heat engines. 
\begin{acknowledgements}
The authors thank M. Hoshina and S. Oono for helpful discussions. 
This study was financially supported by the Hokkaido University
Clark Memorial Foundation.  
\end{acknowledgements}

\bibliography{basename of .bib file}

\end{document}